# Intelligent Search of Correlated Alarms from Database Containing Noise Data

*Qingguo Zheng    Ke Xu    Weifeng Lv    Shilong Ma*
*National Lab. of Software Development Environment*
*Department of computer science and engineering*
*Beijing University of Aeronautics and Astronautics, Beijing 100083, China*
*{zqg,kexu,lwf,slma}@nlsde.buaa.edu.cn*

**Abstract**

Alarm correlation plays an important role in improving the service and reliability in modern telecommunication networks. Most previous research of alarm correlation didn't consider the effects of noise data in the database. This paper focuses on the method of discovering alarm correlation rules from the database containing noise data. We firstly define two parameters Win_freq and Win_add as the measures of noise data and then present the Robust_search algorithm to solve the problem. At different size of Win_freq and Win_add, the experiments on alarm database containing noise data show that the Robust_search Algorithm can discover more rules with the bigger size of Win_add. We also compare two different interestingness measures of confidence and correlation by experiments.

## 1. INTRODUCTION

Telecommunication networks are dynamic, hybrid, heterogeneous and distributed. As a result, it becomes more and more difficult for network administrators to maintain such networks, especially in the fault management, which needs seasoned engineers to do alarm correlation, determine which alarm indicates a fault and find the cause of the fault.

Alarm correlation is a conceptual interpretation of multiple alarms such that new meanings are assigned to these alarms [1]. In the past, the knowledge of alarm correlation was mainly obtained from network experts. But with the development of telecommunication networks, it is much more difficult for experts to keep up with the rapid change of networks. So more and more researchers adopt data mining methods to discover alarm correlation rules.

Data mining is a method of finding interesting patterns from data. It is a generalized inductive learning based on the past cases. The methods of mining

*Accepted by IEEE/IFIP 2002 Network Operations and Management Symposium (NOMS'2002)*, Florence, Italy, April 2002. This research was supported by National 973 Project of China Grant No.G1999032709 and No.G1999032701.

alarm correlation rules are mostly based on the framework of mining association rule algorithm [5] that Agrawal et al. presented in 1993. Mannila and Toivonen et al. [2,3] presented the WINEPI algorithm to find frequent episode from large alarm database, which was applied to the TASA system [4]. Weiss and Hirsh [9] studied how to predict the rare event from alarm database by genetic algorithms and presented the timeweaver algorithm, which was applied to the ANSWER system [10].

Although many methods [1,2,9,20,21,22,23] have been proposed to analyze the alarm correlation, few methods took account of the effects of noise data contained in alarm database. In order to discover alarm correlation rules from alarm database containing noise data, we define two parameters Win_freq and Win_add as the measures of noise data and propose a new algorithm, called Robust_search, which can search the correlated alarms from alarm database containing noise data. At different size of Win_freq and Win_add, experiments on alarm database containing noise data show that the Robust_search Algorithm can discover more rules with the bigger size of Win_add. We also experimentally compare two different interestingness measures of confidence and correlation.

The organization of the paper is as follows. In Section 2, we survey related work and mainly compare our work with Mannila's work. In Section 3, we introduce the problems currently facing alarm correlation and give a definition of *noise data* in this paper. In Section 4, we give a definition of alarm model. In Section 5, we mainly study discovering correlated alarms from the database containing noise and present a new algorithm, called Robust_search, which can discover correlated alarm sequences from alarm database containing noise data. In Section 6, we test the Robust_search algorithm and compare the interestingness measures of correlation and confidence by experiments. In Section 7, we summarize our work and discuss future work.

## 2. RELATED WORK

Agrawal and Srikant [6] first introduced mining sequential patterns from a set of market-basket data sequences, where each sequence element is a set of items purchased in the same transaction. They proposed and experimentally evaluated three algorithms in [6]. Subsequently, Agrawal and Srikant [7] proposed GSP (Generalized Sequential Patterns) Algorithm. The GSP algorithm allows for time–gape constraint, permits that the item of sequence can span a set of transactions within a user-specified window and also permits that the item can span the different item taxonomies.

Mannila et al. [2,3] proposed the WINEPI algorithms to discover the frequent episode from alarm database and classified episode into serial episode and parallel episode. Tuchs and Jobmann adopted Mannila's method to analyze the alarms of GSM networks [13]. Gardner and Harle also adopted the Mannila's method to analyze the alarms of SDH [14].

The differences between our methods and Mannila's methods are explained below. Firstly, we consider the effect of noise data in alarm database and propose the Robust_search algorithm that can search the frequent alarm sequence from alarm database containing noise data, while Mannila et al. didn't consider how to find the episode from alarm database containing noise data. Secondly, we use the number of the times of alarm occurring as the size of windows, while Mannila et al. used the time interval as the size of windows.

Yang and Wang et al. in [15,16] studied mining asynchronous periodic patterns in time series data with noise, proposed a flexible model of asynchronous periodic patterns and a two-phase algorithm. They only considered the serial model in time series data and didn't consider the measure of the periodic patterns in the whole time series data.

# 3. ALARM CORRELATION

## 3.1 The Definition of Alarm Correlation

In network management, a fault is defined as the cause for malfunctioning. An alarm consists of a notification of the occurrence of a specific event, which may or not represent an error [19].

Alarm correlation is converting alarms and merging many alarms into one alarm containing more information. The alarm correlation gives the initial alarm more new meanings [1]. Alarm correlation rules can be used to discover the alarm representing the root cause of fault and exactly locate the fault.

## 3.2 The Problems of Analysis of Alarm Correlation and Definition of Noise Data

Alarms in telecommunication networks are massive, bursting and intermittent. When a fault occurs in the networks, a very large volume of alarms are generated. Network operators are swamped by the alarms, so that it is very difficult for them to discover the root cause of the fault very soon. Thus network operators need a new tool to analyze alarm correlation.

The feature of alarm bursting can be found from figure 1, where the X axis represents the time [mmddhh] and the Y axis represents the number of alarms per hour.

In fact, many alarms don't contain the information about the root cause of fault. When a fault occurs in the networks, the fault may incur many alarms. So some alarms are redundant, which make it more difficult to process the fault. There are some reasons for generating more alarms [17,18].

1. A device may generate several alarms due to a single fault;
2. A fault may be intrinsically intermittent which implies in the sending of a notification at each new occurrence;
3. The fault of a component may result in the sending of an alarm notification

each time the service supplied by this component is invoked;

4. A single fault may be detected by multiple network components, each one of them emitting an alarm notification;
5. The fault of a given component may affect several other components, causing the fault's propagation;
6. There may exist more than two faults at the same time;
7. There is no global network time (no synchronized clock) in the huge and heterogeneous networks. As a result, the sending time stamps of two messages are not exactly comparable.

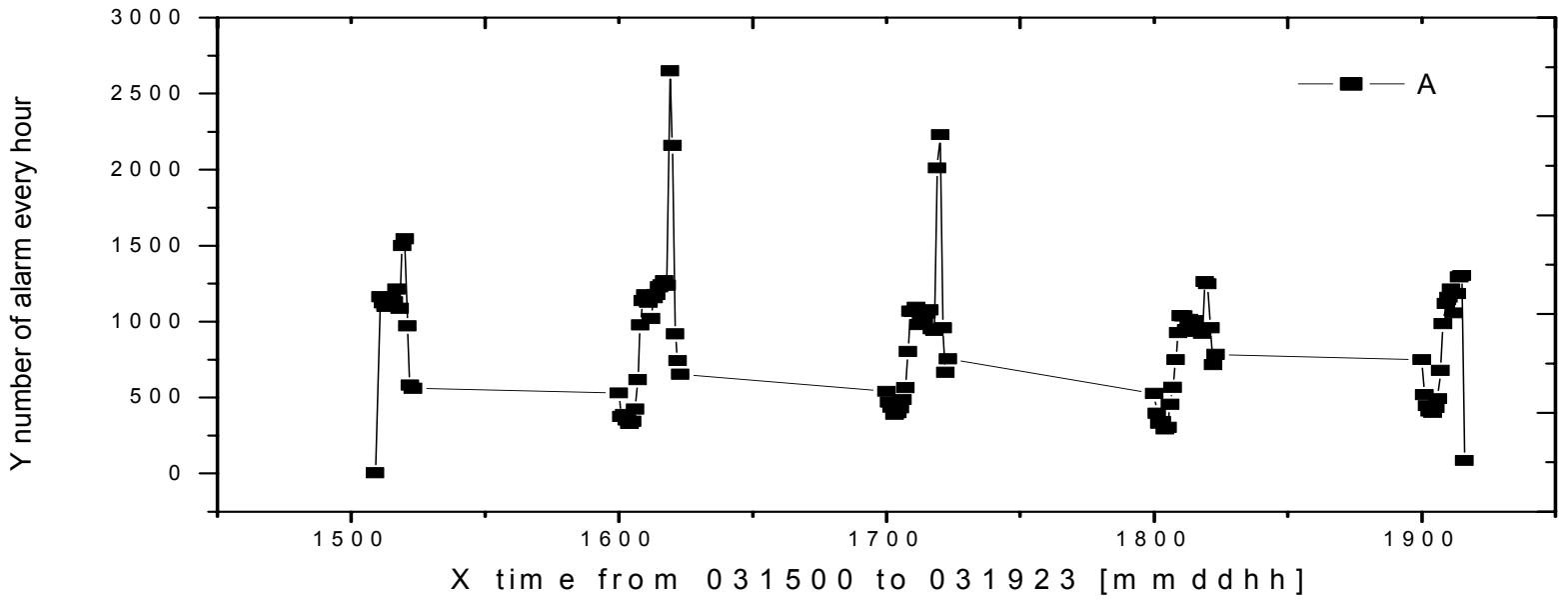

Figure 1:The frequency of alarm occurring

In the analysis of the correlated alarms, we define that the ***noise data*** are alarms that are contained in the correlated alarm sequence, but are not relevant to the correlated alarm sequence. When making analysis of alarm correlation, we must consider the effect of noise data. Otherwise, we can't discover the correlated alarm sequence. In Section 5.2, we will propose a new algorithm, called Robust_search, which can search correlated alarm sequences from alarm database containing noise data.

# 4. ALARM MODEL

An alarm consists of a notification of the occurrence of a specific event, which may or not represent an error [19]. An alarm report is a kind of event report used in the transportation of alarm information.

**Definition 1. An alarm event**

An alarm event is defined as $E_i$=<$e_i$,$t_n$>, i, n=1,2,3,…, where $e_i$ is an alarm type and $t_n$ is the time of alarm occurring.

**Definition 2. An alarm type**

An alarm type is defined as $e_i$=<object_class,object_instance,alarm_num,desc>, i=1,2,3,…, where object_class is the serial NO. of object class, object_instance is the serial NO. of object instance, alarm_number is the NO. of alarm type and desc is the alarm information consisting of alarm priority and alarm description.

**Definition 3. An alarm tuple and its length**

1.) An alarm tuple is defined as $Q_i$=( ( $e_{k1}$, $e_{k2}$, …… , $e_{km}$), $t_i$ ), i,m=1,2,3,…; k1,k2,…,km=1,2,3,… , where '$e_{k1}$, $e_{k2}$,……, $e_{km}$' are the alarm types which concurrently occur at the time $t_i$ . An alarm tuple can be represented by an alarm event i.e. $Q_i$=(< $e_{k1}$, $t_i$>,< $e_{k2}$, $t_i$> ,……, <$e_{km}$, $t_i$>)=( $E_{k1}$, $E_{k2}$, ….. , $E_{km}$). If an alarm tuple $Q_i$ only contains one alarm type, then $Q_i$=(( $e_{k1}$), $t_i$)=( < $e_{k1}$, $t_i$>)= $E_{k1}$.
2.) The length of alarm tuple is $|Q_i|$=|( $e_{k1}$, $e_{k2}$, ……, $e_{km}$)|=m, which is the number of the alarm types contained in the alarm tuple.

**Definition 4. An alarm queue and its length**

3.) An alarm queue is defined as $S_{ij}$=<$Q_i$, $Q_{i+1}$,…, $Q_j$> i,j=1,2,3,…, where $t_i$< $t_{i+1}$ <…<$t_j$ . An alarm queue can be represented by an alarm event i.e. $S_{ij}$=<$Q_i$, $Q_{i+1}$,…,$Q_j$>=< ($E_{i1}$,.., $E_{ik}$), ($E_{i+1}$),…, ($E_{j1}$, …, $E_{jm}$)>= < ($E_{i1}$,…, $E_{ik}$), $E_{i+1}$,…, ($E_{j1}$, …, $E_{jm}$)>, i1,…,ik ,j1,…, jm=1,2,3,… .
4.) The length of alarm queue is defined as | $S_{ij}$|=|<$Q_i$, $Q_{i+1}$, …, $Q_j$>|=j-i+1, which is equal to the number of alarm tuple contained in alarm queue.

**Definition 5. A serial alarm queue and a parallel alarm queue**

Given an alarm queue $S_{ij}$=< $Q_i$, $Q_{i+1}$,…,$Q_j$ >, i, j =1,2,3,⋯ .

1.) If ∀ $Q_k$∈ $S_{ij}$(i≤k≤j), we always have | $Q_i$ |=1, then the alarm queue $S_{ij}$ is called **a serial alarm queue**.
2.) If ∃ $Q_k$∈ $S_{ij}$(i≤k≤j), we have | $Q_i$ |>1, then the alarm queue $S_{ij}$ is called **a parallel alarm queue**.

**Definition 6. An alarm viewing window and its size**

1.) An alarm viewing window is defined as $W_k$=<$Q_m$, …, $Q_n$|d=n-m+1>, where n≥m; k,m,n=1,2,3,⋯ .
2.) The size of alarm viewing window is defined as $|W_k|$=|<$Q_m$,…, $Q_n$|d=n-m+1>|=d , where k,m,n=1,2,3,⋯ .

When making analysis of the correlated alarms, we would rather adopt the length of alarm queues as the size of alarm view windows than the time interval. If we adopt the time interval as the size of alarm view window, for the time intervals of the same length, some may contain a very large number of alarms, while others may contain a very small number of alarms, which will affect the correctness of alarm analysis. Therefore we adopt the times of alarm occurring as the size of alarm viewing window in this paper.

**Definition 7. An alarm type sequence and its length**

1.) An alarm type sequence is the m-tuple consisting of alarm types, which is denoted by $Seq_m$=< $e_{i1}$, $e_{i2}$,…,$e_{im}$> , where m=1,2,3, ⋯ ; i1,⋯,im=1,2,3,⋯ .
2.) Given alarm type sequence $Seq_m$=< $e_{i1}$, $e_{i2}$,…,$e_{im}$>,The length of alarm type sequence is defined as $|Seq_m|$=| < $e_{i1}$, $e_{i2}$,…,$e_{im}$>|=m.

**Definition 8. The time weight of an alarm type sequence, a serial alarm type sequence, and a parallel alarm type sequence**

1.) The time weight of an alarm type sequence is defined as follows. Given an alarm type sequence $Seq_m$=< $e_{i1}$, $e_{i2}$,…,$e_{im}$>, the average intervals among the occurring times of $e_{i1}$, $e_{i2}$,…,$e_{im}$ constitute the time weight of alarm type sequence, which is defined as weight($Seq_m$)=< $\Delta t_1$,$\Delta t_2$,..,$\Delta t_{m-1}$>, m=1,2,3,…; i1…im=1,2,3,⋯.

2.) A serial alarm type sequence is defined as follows. Given an alarm type sequence $Seq_m$=< $e_{i1}$, $e_{i2}$,…,$e_{im}$>, m=1,2,3…, i1, i2,…,im=1,2,3,… and its time weight is weight($Seq_m$)=< $\Delta t_1$, $\Delta t_2$,..,$\Delta t_{m-1}$>. If $\forall \Delta t_k$ ($1 \le k \le m-1$), we always have $\Delta t_k > 0$, k=1,2,3,… , then the alarm type sequence $Seq_m$ is a serial alarm type sequence.

3.) A parallel alarm type sequence is defined as follows. Given an alarm type sequence $Seq_m$=< $e_{i1}$, $e_{i2}$,…,$e_{im}$>, m=1,2,3,…; i1,…,im=1,2,3,… and its time weight is weight($Seq_m$)=< $\Delta t_1$, $\Delta t_2$,..,$\Delta t_{m-1}$>. If $\exists \Delta t_k = 0$ ($1 \le k \le m-1$) k=1,2,3,… , then an alarm type sequence $Seq_m$ is a parallel alarm type sequence.

**Definition 9. The relation of an alarm sequence α is contained in an alarm sequence β**

Given two alarm type sequences α=<$e_{i1}$\`, $e_{i2}$\`,…,$e_{im}$\`>and β=<$e_{i1}$, $e_{i2}$,…,$e_{in}$>, If $n \ge m$, $\exists\ e_{ik} \in \beta$, $\exists\ e_{ij} \in \beta$ ($1 \le ik < ij \le n$) and $e_{i1}$\`= $e_{ik} \cap e_{i2}$\`= $e_{i(k+1)} \cap \ldots \cap e_{im}$\`= $e_{ij}$, then $\alpha \subseteq \beta$.

**Definition 10. An alarm correlation rule**

An alarm correlation rule is defined as

$$e_{i1}, e_{i2}, \ldots, e_{ij} \overset{\Delta t}{\Rightarrow} e_{ik}, e_{ik+1}, \ldots, e_{im} \quad [conf=q\%, supp=p\% ,W_k]$$

After the alarm types $e_{i1}$, $e_{i2}$,…,$e_{ij}$ occur, in the interval of $\Delta t$, the probability of alarm type sequcence <$e_{ik}$, $e_{ik+1}$,…, $e_{im}$> occurring is equal to q% .

# 5. MINING FREQUENT ALARM SEQUENCES

## 5.1 Main Algorithm

To solve the problem that noise data affect the analysis of alarm correlation, we present a new algorithm, called Robust_search (Algorithm 2 described in Section 5.2), which can discover the correlated alarm sequences from alarm database containing noise data.

In this paper, the main algorithm is Algorithm 1, which is based on the framework of mining association algorithm proposed by Agrawal etc al.[5,6,7]. Algorithm 1 is mainly composed of Alogrithm 3 and Algorithm 2. Algorithm 3 generates the alarm type sequence candidates $C_{m+1}$ from frequent alarm type sequence $F_ALARM_m$ and Algorithm 2 counts the times that the alarm type sequence of $C_{m+1}$ occurs in the alarm queue containing noise data.

In what follows, we will introduce the definitions used in Algorithm 1, Algorithm 2 and Algorithm 3.

**Definition 11. Occur($seq_m$, $W_k$), Support($seq_m$, $W_k$), the confidence of the alarm correlation rule and a frequent alarm type sequence**

1.) Given an alarm type sequence $seq_m =< e_{i1}, e_{i2},\ldots,e_{im} >$ and an alarm viewing window $W_k$, the times of the alarm type sequence $seq_m$ occurring in the alarm viewing window $W_k$ are defined as

$$Occur(seq_m,\ W_k)=|\text{ the times of } seq_m \text{ occurring in } W_k\ |$$

2.) Given an alarm type sequence $seq_m =< e_{i1}, e_{i2},\ldots,e_{im} >$ and an alarm viewing window $W_k$, the support of $seq_m$ in $W_k$ is defined as

$$Support(seq_m, W_k)= Occur(seq_m, W_k)/|\ W_k|$$

3.) Given two alarm type sequences: X, Y and an alarm viewing window $W_k$, the confidence of the alarm correlation rule $X\Rightarrow Y$ is defined as

$$Conf(X\Rightarrow Y, W_k)=|Support(XY, W_k) \ / \ Support(X, W_k)- Support(Y, W_k)|$$

4.) A frequent alarm type sequence is defined as follows. In an alarm viewing window $W_k$, given that the minimum support of alarm type sequence is Mini_support, if the support of an alarm type sequence $seq_m$ is greater than Mini_support, then the alarm type sequence $seq_m$ is a frequent alarm type sequence.

Alarm correlation algorithm (Algorithm 1) is composed of two main steps. In the first step, according to the minimum support(Min_support), Algorithm 2 searches the frequent alarm type sequence from alarm queues and the discovered frequent alarm type sequences constitute the set of frequent alarm type sequences, denoted by $F_ALARM_m$. In the second step, according to the confidence of correlation rule, Algorithm 3 generates the alarm correlation rules from $F_ALARM_m$.

■ **Algorithm 1**

Input: alarm queue ($S_{ij}$, $W_k$)

Output: t frequent alarm sequence set: $F_ALARM_m$

```
compute C1:={ α | α∈F_ALARM1};
m:=1;
while  Cm≠Φ  do
 begin
  For all α∈Cm , Search alarm queue Sij to find support(α, Wk); /*Algorithm 2 */
      Obtain  F_ALARMm={ α∈Cm| support(α, Wk)≥ min_support};
      Generate Candidate Cm+1 from F_ALARMm;   /* Algorithm 3 */
      m=m+1;
 end.
for all m , output F_ALARMm;
```

## 5.2 Robust Search Algorithm

If a network facility has a fault, it may incur correlated alarms and other faults may also occur in the network facility and intricate many alarms. So the alarm event sequence that we received will contain many alarms that have nothing to do with alarm correlation. Therefore, when searching the frequent alarm type sequence, we must consider the effects of noise data.

In the followings, we will introduce the definitions in Agorithm 2 and present a new algorithm, called Robust_search(Algorithm 2), which can search the frequent alarm type sequence from the alarm queue containing noise data.

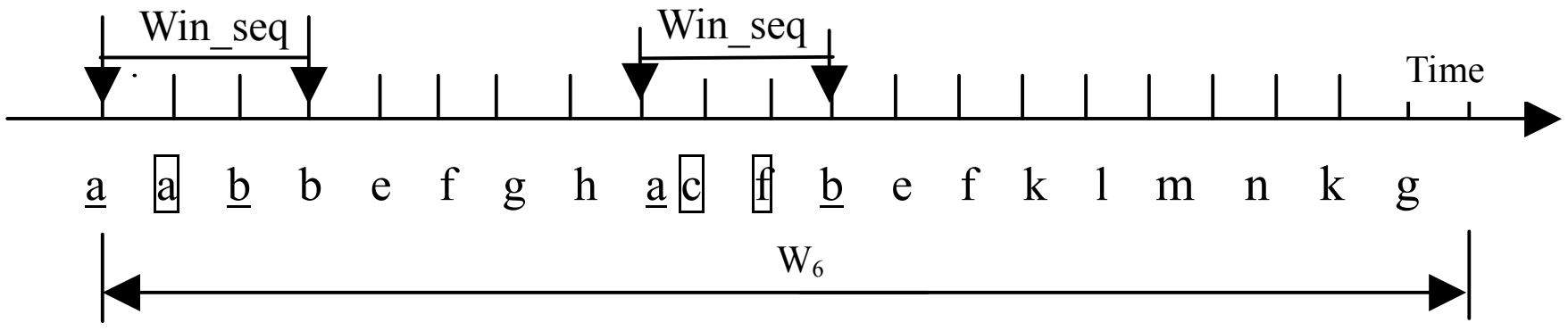

|Win_seq|=4,|Win_freq|=2,|Win_add|=2, $\alpha$=<a,b>, Occur($\alpha$,$W_6$)=2

Figure 2. windows definition in Robust_search Algorithm

**Definition 12. Win_seq, Win_freq and Win_add**

1.) An alarm search window, denoted by Win_seq, is defined as the scope of the alarm type sequence being searched in the alarm queue.
2.) The size of alarm search window is defined as |Win_seq|=|<$E_m$,…,$E_n$>|=n-m+1, where |Win_seq|=|win_freq|+|win_add|.
3.) A frequent alarm window, denoted by Win_freq, is defined as the base window of the alarm type sequence of search.
4.) Given an alarm sequence $\alpha$=< $e_{i1}$, $e_{i2}$,…,$e_{im}$ > and $\alpha \in C_m$ where $C_m$ are composed of the alarm type sequence candidates of length m, The size of a frequent alarm window is defined as |Win_freq|=|$\alpha$|$\alpha \in C_m$|=|< $e_{i1}$, $e_{i2}$,…,$e_{im}$ >|=m, which is equal to the length of the alarm type sequence of search.
5.) An additional alarm window, denoted by Win_add, is defined as the noise data tolerance window to search the alarm type sequence in an alarm search window Win_seq.
6.) The size of additional alarm window is defined as the maximum number of noise alarms included in the alarm search window Win_seq. We have a alarm sequence containing noise alarms $\beta$=<$e_{i1}$, $\boxed{e_{j1}}$,$e_{i2}$, $e_{i3}$,$\boxed{e_{j2}}$,…,$\boxed{e_{jk}}$,…,$e_{im}$ > in Win_seq, where an alarm type sequence in $\beta$ is < $e_{i1}$, $e_{i2}$,…,$e_{im}$ > and the noise alarm type sequence in $\beta$ is < $\boxed{e_{j1}}$,$\boxed{e_{j2}}$,…,$\boxed{e_{jk}}$>. The length of the additional alarm window is |Win_add|=| < $\boxed{e_{j1}}$,$\boxed{e_{j2}}$,…,$\boxed{e_{jk}}$>|=k, as well as |Win_freq|=|< $e_{i1}$, $e_{i2}$,…,$e_{im}$ >|=m and |Win_seq|=k+m.

Robust_search(Alogrithm 2.) is mainly described as follows. Given an alarm queue $S_{n1\,nq}$ =<$Q_{n1}$, $Q_{n2}$,……,$Q_{nq}$> in the alarm viewing window $W_k$, in order to explain more clearly, we will adopt the alarm event to represent the alarm queue ,i.e. $S_{n1\,nq}$ =<($E_{n1\,0}$, …, $E_{n1\,j1}$),……, ($E_{nq\,0}$, …, $E_{nq\,jq}$)>. Given an alarm type sequence $\alpha$=< $e_{i1}$, $e_{i2}$,…,$e_{im}$>, the size of window i.e. Win_seq is equal to m+|Win_add|. At the

beginning, the pointer Ptr_seq point to $E_{x1\ y1}$ (x1,y1=1,2,3,…)(5th line in Algorithm 2.), which is the first alarm event containing the alarm type '$e_{i1}$' in $W_k$. From $E_{x1\ y1}$ to the end of the alarm search window Win_seq, Algorithm 2 looks for the alarm event containing the alarm type '$e_{i2}$'. if there exists $E_{x2\ y2}$ (x2,y2=1,2,3,...) containing alarm type $e_{i2}$, then after the alarm event $E_{x2\ y2}$, Algorithm 2 goes on to look for the alarm event containing alarm type $e_{i3}$, and so on. If $e_{i1}, e_{i2},\ldots,e_{im}$ are all found in the scope of alarm search window Win_seq, then we say that the alarm type sequence $\alpha=< e_{i1}, e_{i2},\ldots,e_{im}>$ occurs one time in $W_k$. After that the pointer Ptr_seq is moved to the next alarm event containing '$e_{i1}$' behind the alarm event $E_{xm\ ym}$(xm,ym=1,2,3, …) that matches the last alarm type $e_{im}$ of $\alpha$ in the previous search; If the alarm sequence $\alpha=< e_{i1}, e_{i2},\ldots,e_{im}>$ is not contained in the alarm search window Win_seq, then the start pointer Ptr_seq is moved to the next alarm event containing '$e_{i1}$'. After finishing the search in the alarm queue $<Q_{n1}, Q_{n2},\ldots\ldots,Q_{nk}>$, we will obtain occur($\alpha$,$W_k$) (19th line in Alogrithm2).

An example about the Robust_search algorithm is illustrated in figure 2.

■ **Algorithm 2**

Input: frequent alarm Candidate $C_{m,}$ $W_k$ , win_add

Output: Occurr ($C_m$, $W_k$)

1. |Win_seq|:=m+|Win_add|; /* |Win_freq|=m */
2. for ( ; $\alpha \in C_m$ ; ) /* $\alpha=< e_{I1}, e_{I2},\ldots,e_{im}>$, $W_k=<Q_{n1}, Q_{n2},\ldots\ldots,Q_{nq}|d=q>$ */
3. begin /* C_count keeps the Occur($\alpha$,$W_k$)*/
4. C_count:=0; /* Ptr_seq is the start pointer of Win_seq*/
5. Ptr_seq point to the first alarm event containing '$e_{i1}$';
6. while((Ptr_seq+m) is in $W_k$ )
7. begin
8. Ptr_temp:=Ptr_seq
9. for(p:=1;p<=m;p++)
10. if(from Ptr_temp to the end of Win_seq, '$e_{i\ p}$' is not found) then break;
11. else Ptr_temp point to the next alarm event in Win_seq;
12. if(p=m+1) begin
13. The $\alpha$ occurs one time in Win_seq, C_count++;
14. Ptr_seq points to the first alarm event containing alarm type '$e_{i1}$'
15. after the alarm event that Ptr_temp points to;
16. end
17. else Ptr_seq points to the next alarm event containing '$e_{i1}$';
18. end
19. Occur($\alpha$,$W_k$):=C_count;
20. end /* end of for loop */

## 5.3 The Complexity Analysis of Robust_search Algorithm

Given an alarm viewing window $W_k$ where $|W_k|$=d, we have an alarm event queue $S_{1d}=< Q_1, Q_2, \ldots\ldots, Q_d>$, an alarm type sequence $\alpha=< e_{i1}, e_{i2}, \ldots, e_{im}>$ and an alarm search window Win_seq where |Win_seq|=m+|Win_add|. Let Supp=Support($<e_{i1}>$,

$W_k$) and M=Max{ $|Q_i|, 1 \le i \le d$}. Then the worst time complexity of Robust_search algorithm on $W_k$ is M ·d ·(1+Supp · |Win_seq|/M).

**Proof:**

In the Robust_search algorithm, the alarm search window slides ($|Q_1|+|Q_2|, \ldots, +|Q_d|) \le M \cdot d$ times on alarm queue $S_{1d}$ in alarm viewing window $W_k$. Note that Supp=Support(<$e_{i1}$>, $W_k$), so there are at most d·Supp alarm events containing '$e_{i1}$'. At each alarm event containing '$e_{i1}$', the algorithm will do matching for |Win_seq| times at most. Then the algorithm will do d·Supp·|Win_seq| times matching in the worst case. Therefore the worst time complexity of Robust_search algorithm is M·d·(1+Supp·|Win_seq|/M).

In the experiments of this paper, M is about 10, Supp is about 0.002 and |Win_seq| is about 10, then the time complexity of Robust_search algorithm mainly depends on the number of alarm events in $W_k$. Since |Win_seq|=m+|Win_add| and the value of Supp is very small in general, the size of Win_add has little effect on the time complexity of Robust_search algorithm.

## 5.4 Generate Candidate Agorithm

Algorithm 3 is composed of two steps. In the first step, the algorithm generates alarm type sequence $\gamma$=< $e_{i1}$, $e_{i2}$, …, $e_{im}$, $e_{im}$`>,where | $\gamma$ |=m+1, from $F_ALARM_m$ . In the second step, if all the subsets L of the alarm type sequence $\gamma$, where |L|=m, are contained in $F_ALARM_m$, then $\gamma$ belongs to $C_{m+1}$.

- **Agorithm 3**

Input: frequent alarm sequence set $F_ALARM_m$.

Output: frequent alarm sequence Candidate set $C_{m+1}$.

1. $C_{m+1}:=\Phi$;
2. For $\alpha,\beta \in F_ALARM_m$ and $\alpha \ne \beta$ and $\alpha$=< $e_{i1}$, $e_{i2}$,…, $e_{im}$>, $\beta$=< $e_{i1}$\`, $e_{i2}$\`,…, $e_{im}$\`>
3. begin
4. If ($e_{i2}$= $e_{i1}$\` $\cap e_{i3}$= $e_{i2}$\` $\cap$,…,$\cap e_{im}$= $e_{i\,m-1}$\` ) then begin
5. generate alarm sequence $\gamma$=< $e_{i1}$, $e_{i2}$,…, $e_{im}$, $e_{im}$\`>; /* Candidate generate */
6. $C_{m+1}$:= {$\gamma$ | For all $L \subseteq \gamma$ and |L|=m ,we have $L \in$ $F_ALARM_m$ };/*Pruning Candidate */
7. end
8. end

## 5.5 Generate Correlation Rules

There are various interestingness measurers of rules in the methods of data mining. R.Agrawal et al. in [5] first presented AIS association rules algorithm and its measure of association rule $X \Rightarrow Y$, which is defined as confidence($X \Rightarrow Y$) =Support(XY)/Support(X), where X and Y correspond to a set of attributes and X and Y are disjoint.

Brin et al. [11] studied generalizing Association Rules to Correlation as follows. The support and confidence of an association rule $X \Rightarrow Y$ are defined as Support=P[XY] and Confidence=P[XY]/P[X]. The confidence is the conditional probability of Y given X. If X and Y are independent, then Confidence =P[XY]/P[X]=P[Y]. Therefore, if P[Y] is high, then the confidence of the rules is high, which will make association rule meaningless. In order to solve the problem, S.Brin et al. [11] discussed measuring significance of association rules via the support and the chi-quared test for correlation and they also presented the interestingness measure $I=P(XY)/(P(X)\times P(Y))$. The interestingness measure is symmetrical, because the confidence of $X\Rightarrow Y$ is equal to the one of $Y\Rightarrow X$. However, Khailil M et al.[12] proved that the chi-squared test is correct for 2×2 continuos tables, but incorrect for the larger continuous table and they also presented the new interestingness measure $R(X\Rightarrow Y)=|P(XY)/P(X)-P(Y)|$. In this paper, we experimentally compare $R(X\Rightarrow Y)=P(XY)/P(X)-P(Y)$ with Confidence =P[XY]/P[X], which is described in detail in experiment 2.

In [5] the association rules only have one single item of the consequent, then R. Agrawal and R. Srikant in [8] gave an algorithm of generating more than one item of the consequent. The Algorithm 4 is based on the algorithm generating association rules in [8] and the interestingness measure of the correlation rules adopts $R(X\Rightarrow Y)=P(XY)/P(X)-P(Y)$ in [11]. A rule holds if and only if the confidence of rule is greater than min_conf.

■ Algorithm 4

```
Input:   Frequent alarm sequence set F_ALARM_m
Output:  output the correlation rules β→(α-β) and
         confidence  |P(α)/P(β)-P(α-β)|
for all α∈F_ALARM_m do /* generate  correlation  rules */
    for all β⊆α do
        if|P(α)/P(β)-P(α-β)|≥min_conf  then
            begin
            generate the rule β→ (α-β) with
                confidence |P(α)/P(β)-P(α-β)| ;
            end
```

# 6. EXPERIMENTS

## 6.1 The Results of Experiment 1

The data in experiment 1 are the alarms in GSM Networks, which contain 181 alarm types and 91311 alarm events. The time of alarm events ranges from 2001-03-15-00 to 2001-03-19-23. In figure 3 the broken line graph is denoted by win_xy, where x represents the size of additional alarm window i.e.Win_add and y represents the size of frequent alarm window i.e. Win_freq. In figure 3 the Y axis is the number of alarm type sequences and the X axis is Mini_support (using the minimum occurring times ).

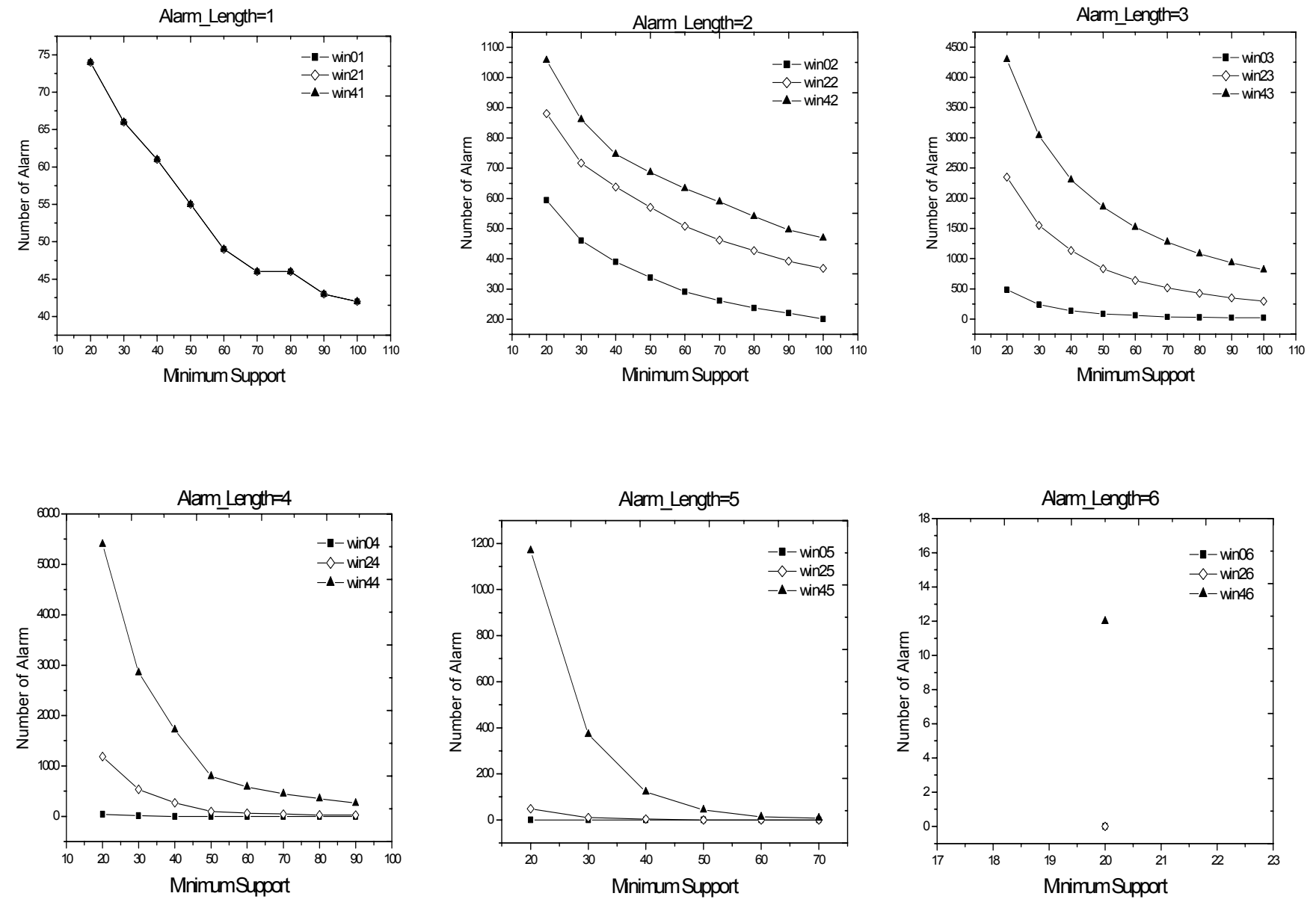

Figure 3:The number of frequent sequences changes with Win_add

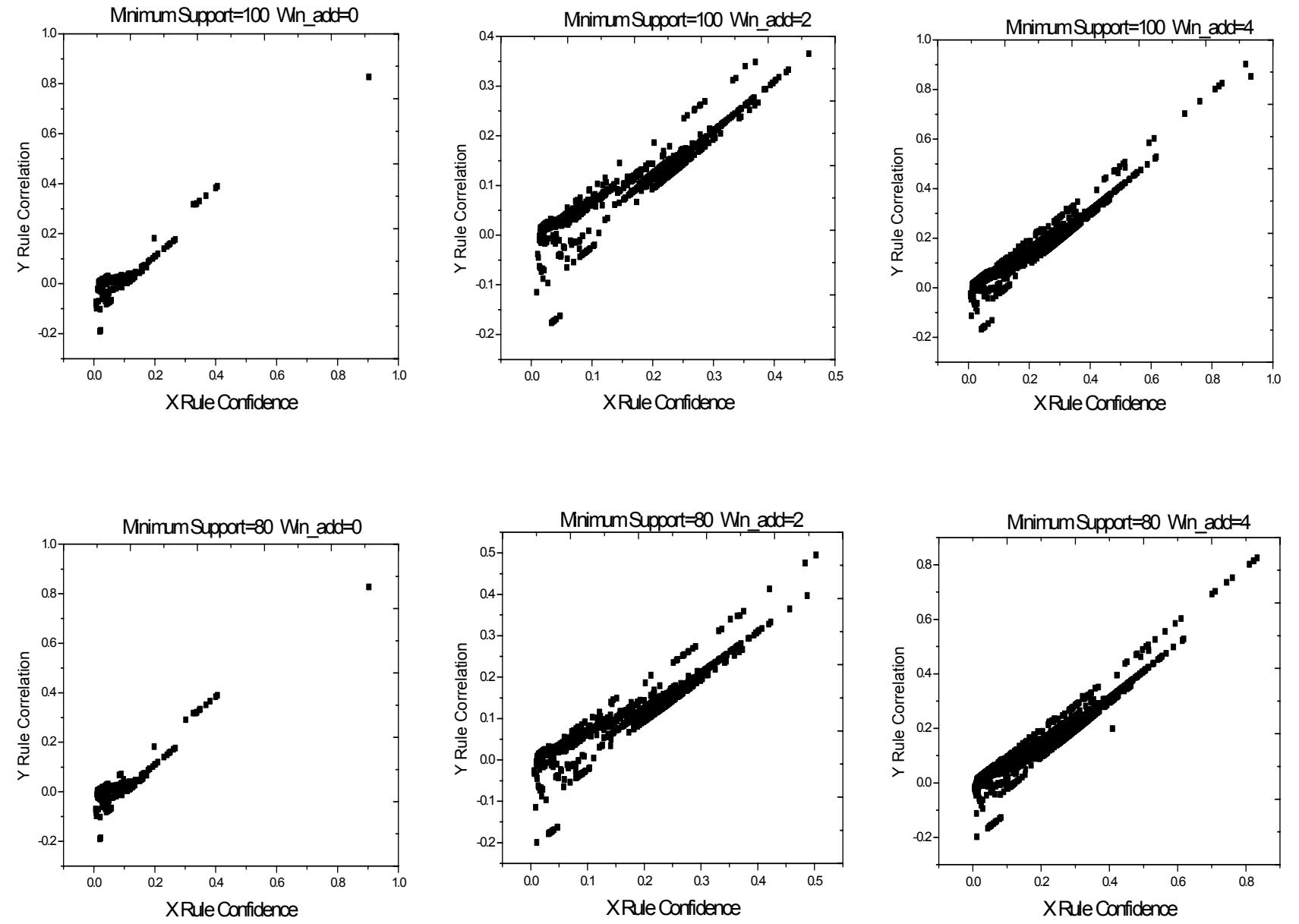

Figure 4: Interestingness measure: Correlation and Confidence

The results of experiment 1 are illustrated in figure 3. From figure 3, it is also easy to see that as y increases, win_0y, win_2y and win_4y (y=1,2,3,4,5) become close to each other more rapidly with the increment of Mini_support.

In sum, with the increment of the size of additional alarm windows, the number of frequent alarm type sequences increases and the number of alarm sequences will increase more slowly with the increment of the length of alarm sequence.

### 6.2 The Results of Experiment 2

The data in experiment 2 are the same as those in experiment 1. In figure 4 the X axis represents Confidence=P[XY]/P[X] and the Y axis represents Correlation =P(XY)/P(X)-P(Y).

The results of experiment 2 are illustrated in figure 4. The interestingness measures of correlation rules are mainly distributed in [0, 0.4], especially in[0, 0.1]. In figure 4, at the same support, with the increment of the additional alarm window i.e. Win_add, the number of correlation rules will increase. As Win_add increases, the increment of high confidence rules is greater than that of lower ones. The relation between correlation and confidence is nearly linear and its slope is lower than 1.

## 7. CONCLUSION

Since the Robust_search algorithm can analyze alarm correlation from alarm database containing noise data, it will generate more alarm sequences, then the number of correlation rules increases. Although the correlation measure can reduce the rules, it still needs people to select the most useful ones from a large number of the rules. Therefore, it is necessary to study how to extract rules more correlated from alarm database containing noise in the future.

**ACKNOWLEDGMENTS**

This research was supported by National 973 Project of China Grant No.G1999032709 and No.G1999032701. Thanks Professor Wei Li for the choice of subject and guidance of methodology. Thanks for the suggestions from Professor Yuefei Sui of the Institute of Computing Technology, Chinese Academy of Sciences. Thanks for Nan Liu, Zhi Cui and Xinzhang Li of Beijing Mobile Telecom. The authors would also like to thank the other members of the NLSDE lab: Xin Gao, Peng Cheng, Gang Zhou, Lin Tian, and Dong Li.

# References

[1] G. Jakobson and M. D. Weissman, “Alarm Correlation,” IEEE Network, page52-59, November 1993.

[2] H. Mannila and H. Toivonen and A. I. Verkamo, “Discovering frequent episodes in sequences,” KDD-95, pp. 210-215, August20-21, 1995.

[3] H. Mannila and H. Toivonen, “Discovering Generalized Episodes Using

Minimal Occurrences," *Proceedings of the seconds International Conference on Knowledge Discovery and Data Mining*, Portland, Oregon, August 1996.

[4] K. Hät\"onen, M. Klemettinen, H. Mannila, P. Ronkainen, and H. Toivonen, "TASA: Telecommunications Alarm Sequence Analyzer, or How to enjoy faults in your network," *IEEE/IFIP 1996 Network Operations and Management Symposium (NOMS'96)*, Kyoto, Japan, April 1996, p. 520-529.

[5] R. Agrawal, T. Imielinski, and A.N. Swami, "Mining Association Rules Between Sets of Items in Large Databases," *Proceedings of the 1993 ACM SIGMOD International Conference on management of data*, pp. 207-216, Washington, D.C., May 1993.

[6] R. Agrawal and R. Srikant, "Mining Sequential Patters," *In Proceedings of the Eleventh International Conference on Data Engineering*, Taipei, Taiwan, March 1995.

[7] R. Srikant and R. Agrawal, "Mining Sequential Patterns: Generalizations and Performance Improvements," *In Proceedings of the Fifth International Conference on Extednding Database Technology (EDBT'96)*, Avigonon, France, March 1996.

[8] R. Argawal and R. Srikant, "Fast Algorithms for Mining Association Rules," *In Proceedings of the 20$^{th}$ international Conference on Very Large DataBase*, Santiago, Chile, September 1994.

[9] G. M. Weiss and H. Hirsh, "Learning to Predict Rare Events in Event Sequences," *In proceeding 4th International Conference on Knowledge Discovery and Data Mining*, AAAI press, 1998, pp.359-363.

[10] G. Weiss, J. Eddy, S. Weiss, "INTELLIGENT TELECOMMUNICATION TECHNOLOGIES," Knowledge-Based Intelligent Techniques in Industry, pp.249-275, 1998.

[11] S. Brin, R. Motwani and C. Silverstein, "Beyond Market Baskets:Generalizing Association Rules to Correlations," *In The Proceedings of SIGMOD*, pp. 265-276, AZ, USA, 1997.

[12] K. M. Ahmed, N. M. EI-Makky, Y. Taha, "A note on 'Beyond Market Baskets:Generalizing Association Rules to Correlations'," SIGKDD Explorations, Volume1, Issue2-page46, Jan, 2000.

[13] K. D. Tuchs, K. Jobman, "Intelligent Search for Correlated Alarm Events in Database", *IFIP/IEEE International Symposium on Integrated Network Management*, 14-18 May 2001.

[14] R. D. Gardner and D. A. Harle, "Fault Resolution and Alarm Correlation in High-Speed Networks using Database Mining Techniques," *Information, Communications and Signal Processing 1997*, pp.1423-1427, September 1997.

[15] J. Yang, W. Wang, and P. Yu. "Mining asynchronous periodic patterns in time series data," *In The Proc. ACM SIGKDD Int. Conf.* O*n Knowledge Discovery and Data Mining (SIGKDD)*, pp.275-279, 2000.

[16] W. Wang, J. Yang, P. S. Yu, "Mining Patterns in Long Sequential Data with

Noise," ACM SIGKDD, Volume2, Issue2, pp.28-33, December 2000.

[17] K. Houck, S. Calo, and A. Finkel, "Towards a practical alarm correalation system," *In IFIP/ IEEE International Symposium on Integrated Network Management,IV(ISINM'95)*, pp. 226-237.

[18] Lehmann, Diesel, Deters, Seibold, Sevcik, Uhde, Schmid, Huber, Muetter-Feldbusch, Syrjakow, Szczerbicka, Ziegler, "Knowledge-Based Alarm Surveillance for TMN," *In Proceedings of the International IEEE Conference on Computer and Communications*, 1996, pp. 494-500.

[19] ITU-T. "Recommendation X.720:Information technology-Open Systems Interconnection-structure of management information: Management information model," January 1992.

[20] A. T. Bouloustas, S. B. Calo, A. Finkel, "Alarm Correlation and Fault Identification in Communication Networks," IEEE Transactions on Communications, Vol.42, No.2/3/4, Feb/Mar/Apr 1994.

[21] I. Katzela, A. T. Bouloutas, and S. B. Calo. "Centralized vs. distributed fault localization," *In IFIP/IEEE International Symposium on Integrated Network Management*, pp 250-261.

[22] S. Kliger, S. Yemini. "A Coding Approach to Event Correlation," *Integrated Network Management proceedings of the fourth international symposium on integrated network management 1995*, pp.266-277, Chapman & Hall, ISBN 0-412-71570-8, 1995.

[23] L. Lewis, "A Case-Based Reasoning Approach to resolution of faults in communication networks," *In Integrated Network Management*, III, 671-682, Elsevier Science Publisher B.V., Amsterdam., 1993

## Biograhies

**Qingguo Zheng** received his MS in Computer Science and Engineering from JiLin University in 1998. Currently he is working towards his PHD in Computer Science at *National Lab. of Software Development Environment,* Beijing University of Aeronautics & Astronautics. His research interests include Network Management, Data Mining. His permanent e-mail: qingguozheng@263.net

**Ke Xu** received his PHD in Computer Science from Beijing university of Aero. and Astro. in 2000. Now he is a Postdoc in *National Lab. of Software Development Environment*. His research interests include the design and analysis of algorithms and computational complexity. His permanent e-mail: ke.xu@263.net

**Weifeng Lv** is an Associate Professor in Department of computer science and engineering, Beijing university of Aero· and Astro·. His research interests include the SAT problem, Network Management.

**Shilong Ma** is a Professor in Department of computer science and engineering, Beijing university of Aero· and Astro·. His research interests include Logic and Symbolic Computation.